# TREBLE: Fast Software Updates by Creating an Equilibrium in an Active Software Ecosystem of Globally Distributed Stakeholders

KEUN SOO YIM, Google
ILIYAN MALCHEV, Google
ANDREW HSIEH, Google
DAVE BURKE, Google

This paper presents our experience with TREBLE, a two-year initiative to build the modular base in Android, a Java-based mobile platform running on the Linux kernel. Our TREBLE architecture splits the hardware independent core framework written in Java from the hardware dependent vendor implementations (e.g., user space device drivers, vendor native libraries, and kernel written in C/C++). Cross-layer communications between them are done via versioned, stable inter-process communication interfaces whose backward compatibility is tested by using two API compliance suites. Based on this architecture, we repackage the key Android software components that suffered from crucial post-launch security bugs as separate images. That not only enables separate ownerships but also independent updates of each image by interested ecosystem entities. We discuss our experience of delivering TREBLE architectural changes to silicon vendors and device makers using a yearly release model. Our experiments and industry rollouts support our hypothesis that giving more freedom to all ecosystem entities and creating an equilibrium are a transformation necessary to further scale the world largest open ecosystem with over two billion active devices.

## 1 INTRODUCTION

Android uses ecosystem- and supply chain-based operating system (OS) development model. The framework vendor (i.e., Google) publishes new versions of the Android platform on Android Open Source Project (AOSP). Silicon vendors in the world then customize AOSP for their chipsets (e.g.,

Author's addresses: K. S. Yim, Google LLC, Attn yim, 1600 Amphitheatre Parkway, Mountain View, CA 94043, USA; I. Malchev, Google LLC, Attn malchev, 1600 Amphitheatre Parkway, Mountain View, CA 94043, USA; A. Hsieh, Google LLC, Attn andrewhsieh, 1600 Amphitheatre Parkway, Mountain View, CA 94043, USA; D. Burke, Google LLC, Attn daveburke, 1600 Amphitheatre Parkway, Mountain View, CA 94043, USA.

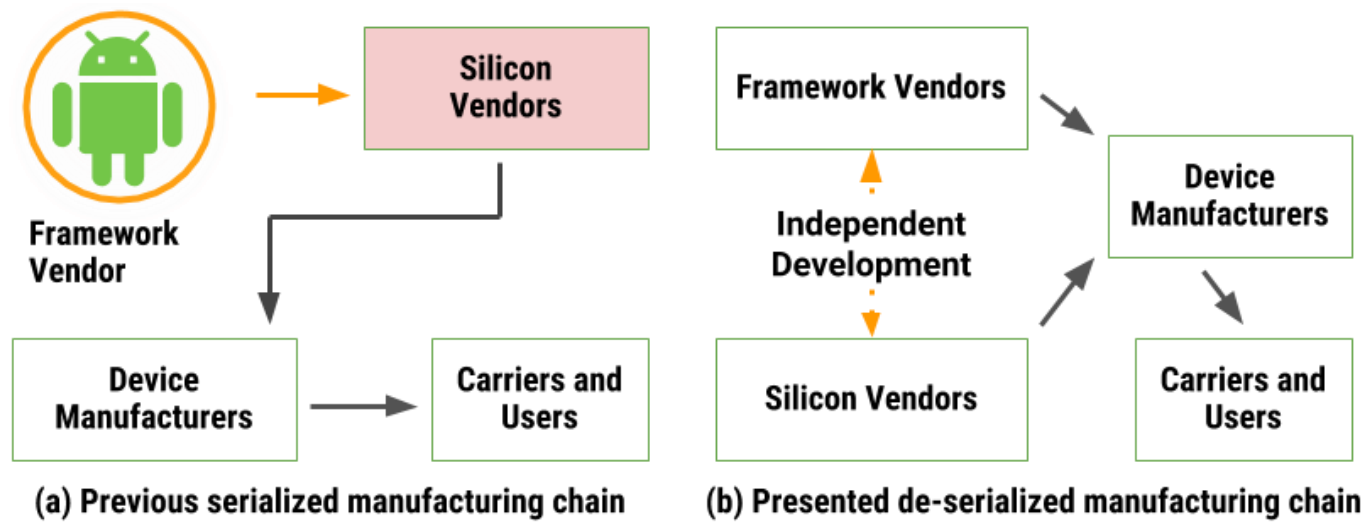

Fig. 1. Android device manufacturing ecosystem models.

Table 1. Android Ecosystem Updatability Dilemma

| Ecosystem Entity | Desire to Launch | Desire to Update | Desire to Extend |
|---|---|---|---|
| **Framework vendors** | High | High | Low |
| **Silicon vendors** | High | Low | Medium |
| **Device manufacturers** | High | Medium | High |
| **Carriers and users** | Medium | Medium | Medium |

for system-on-chip-specific enhancements and fixes). Such Android extensions (namely, *AOSP forks*) become the source of truth for Android device makers. Large manufacturers further customize those extensions heavily (e.g., with special features and value-added applications) as depicted in Figure 1(a). In this way, device makers, whose core strengths are in system integration and mass production, could consistently use a royalty-free, competitive OS whose rapid growth makes them now part of the world largest device ecosystem for mobile applications.

The large and growing deltas between stock Android and forks created by silicon vendors and device makers pose vital challenges in maintaining the software updatability [1][2][3] crucial for the security, quality, and thus overall health of the Android ecosystem. This *platform updatability challenge* originates with silicon vendors, who are interested in fast initial device launches, but are less motivated to maintain the software of previous generation chipsets (see Table 1). However, other ecosystem entities still want to get fast platform software updates. In the previous serialized Android manufacturing chain model shown in Figure 1(a), the framework vendor needed to seek out close collaborations with silicon vendors and device makers. That collaboration was difficult because most silicon vendors are not highly interested in, making it difficult to update even the framework part of the Android platform on existing Android devices.

Device manufacturers also have difficulties in taking a new Android framework release if they have significantly extended a previous Android framework, and the previous Android platform does not have an extensible architecture. That is typically the case because as shown in the extension column of Table 1, device manufacturers are the most interested in framework extensions. A more fundamental problem is that if a new platform is developed for a new system-on-chip (SoC), then device manufacturers need to pay a significant engineering cost to adapt that new platform to older (but likely the most widely used) Android devices that use previous system-on-chips. The resulting *framework version fragmentation problem* could expose a large number of Android devices to known security vulnerabilities.

To address the framework forking and fragmentation problems, in this study, we transform how components in the Android platform software stack are supplied to other entities in the ecosystem. As depicted in Figure 1(b), our deserialized manufacturing model allows individual entities to own certain parts of the Android platform stack and perform independent software updates. The basis for that is our TREBLE architecture, which redefines *Android framework*. In our new definition, every system layer component that is system-on-chip- and device-agnostic is part of the Android framework. Code that does not know what hardware it is running on or compiled for is part of the framework. The framework includes Java code along with native[1] libraries (e.g., libc, libm, libstdc++). It also includes, conceptually, those parts of the kernel agnostic of system-on-chip or device on which they are running.

The core of TREBLE re-architecting is thus defining the new vendor interface (VINTF) to separate silicon- and device-specific code (namely, *vendor interface implementation*) from the core OS. The redefined Android framework communicates with hardware in several different ways. Thus, the vendor interface API surface straddles several boundaries such as the hardware abstraction layer (HAL)[2], vendor-specific native libraries, and the Linux kernel. The focus of vendor interface design is on formally defining and versioning the APIs for silicon- and device-dependent code, and fully decoupling the vendor interface implementation below the vendor interface from the Android framework. The vendor interface is enforced by the *Vendor interface Test Suite (VTS)*.

We then repackage the Android framework as a separate system image. The redefined system.img logically contains much more than in previous Android framework images. For example, native libraries that code in vendor interface implementation depends on[3] are now part of system.img. This allows framework vendors and device manufacturers to update such libraries by themselves if they fix bugs or vulnerabilities. Similarly, we adopt dynamically loadable kernel modules (DLKMs) and repackage kernel modules as part of vendor.img, while optionally migrating the kernel itself to boot.img. This allows silicon vendors who work closely with the upstream Linux community to update the most vulnerable parts of the kernel (i.e., device drivers especially closed source binary drivers) as part of the vendor.img update, which lies under their control when using the TREBLE architecture. We also include various automated tests (e.g., greybox fuzzing) in vendor interface test suite so that Android device partners can discover vulnerabilities and other bugs in vendor.img before shipping their devices. This reduces the need for vendor.img security updates.

It is challenging to transform the architecture and rewrite the existing components of an active software platform ecosystem of various globally distributed stakeholders – the cost of transformation grows the longer it takes. This paper includes the four principles we used for re-architecting the Android platform stack and performing the transformation between 2016 and 2018. This paper also shares our experience in: (1) building various system integration and productivity tools and infrastructure; (2) working with more than 30 development teams for rewriting and test development of various components; and (3) continuously sharing the new changes to key partners in the ecosystem while handling component development and integration. Our developments during the two year period of industry wide rollouts show the improvements in Android platform updatability. These also help us experimentally evaluate the engineering cost of re-architecting, and how the associated performance and power overheads are managed.

The rest of this paper is organized as follows. Section 2 describes the background and challenges. Section 3 presents our engineering principles. Section describes our TREBLE design. Section 5 analyzes our experimental results. Section 6 discusses our experience in integration and deployment and summarizes the lessons learned. Section 7 reviews related works. Section 8 concludes.

## 2 BACKGROUND

[1] Native refers to C/C++ code because Android is written in Java.

[2] A HAL is logically equivalent to a user space device driver wrapping a respective kernel driver (e.g., with ioctl() calls). It is part of a vendor implementation but not part of the Android framework.

[3] C/C++ native libraries that are used solely by a vendor interface implementation (e.g., by HALs). TREBLE makes most of them (namely, Vendor NDK) as part of the Android framework.

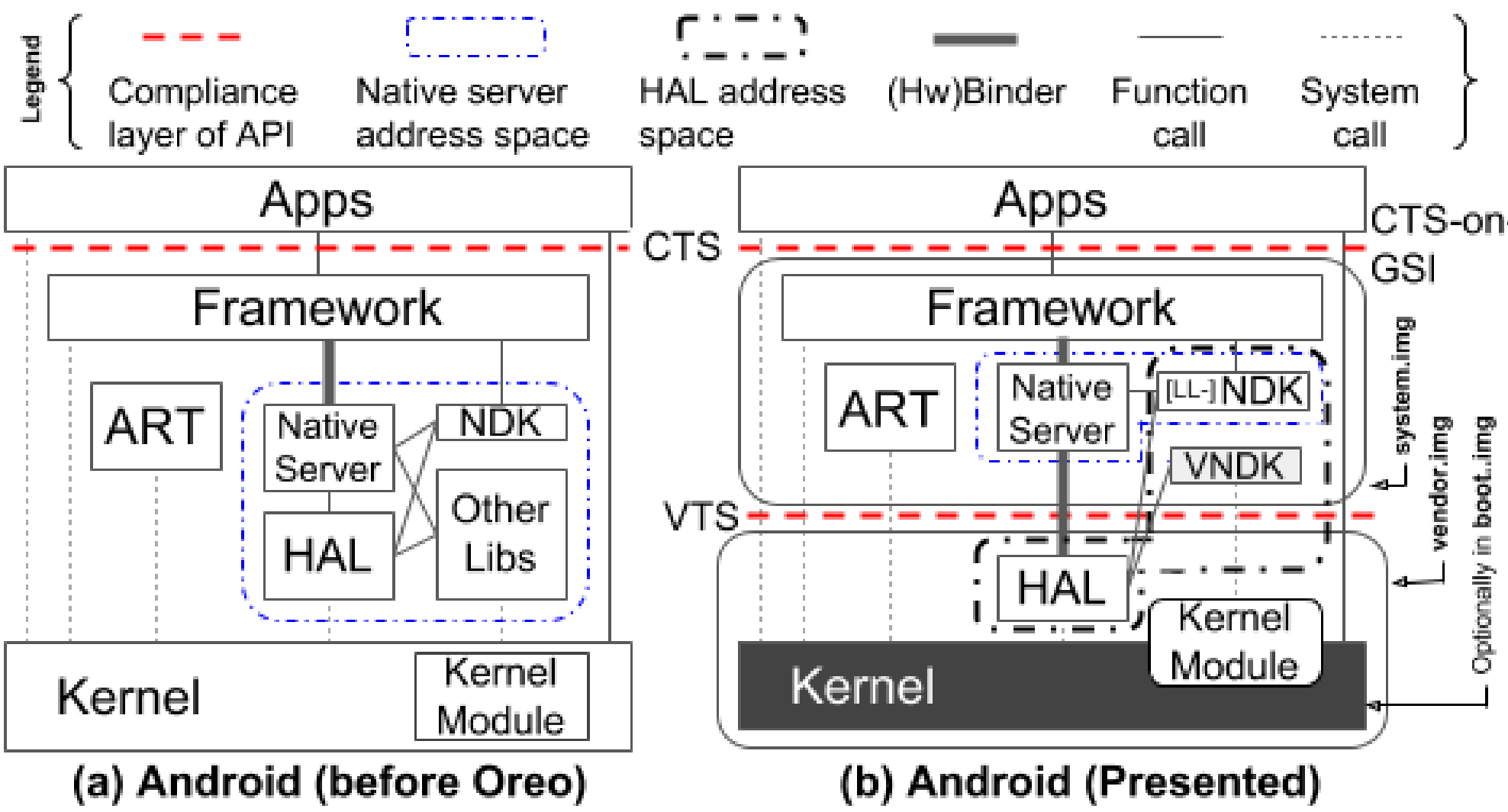

Fig. 2. Modular Android platform software architecture.

This section describes the architecture of the previous Android software platform and the challenges in modularizing its components.

We first focus on management models for app and platform API compatibility. Figure 2(a) represents what Android has used before project TREBLE. The framework and the kernel were tightly integrated; no formally-defined and enforced API was necessary and thus none exists between them. However, the previous Android model explicitly defines an API layer between apps and the framework in its compatibility definition document and enforces that through the compatibility test suite (CTS) [5]. Our model uses a new API compliance layer (see Figure 2(b)) between the framework and the kernel, and simultaneously enforces those two API layers to achieve the fore described design objectives.

Let us analyze the components used in the previous Android platform software stack. As shown in Figure 2(a), the Android framework in the center is Java-based middleware running on top of Android Runtime (ART), which executes Dalvik executables (.dex bytecodes). The framework provides a rich set of APIs to Android apps through control of and interaction with various native servers (e.g., using an inter-process communication mechanism), native libraries (aka Native Development Kit, NDK)[4], and the Linux kernel. Native servers provide various subsystem services (e.g., camera, audio, and media) and are implemented as independent Linux user processes.

Binder [4] is an inter-process communication (IPC) mechanism implemented by using its Linux kernel module and user space libraries. Binder supports both Java and C/C++ clients and servers. Android Interface Definition Language (AIDL) is typically used to specify binder server APIs because an AIDL tool can then automatically generate both server stub and client proxy code from a given .aidl specification file. A native server typically used one or multiple conventional hardware abstraction layer (HAL) modules in its address space (see Figure 2(a), address space box). A conventional HAL was a shared library with a special symbol for its API lookup.

Both the framework and HALs can use native libraries. Through the Java native interface (JNI), the framework uses native libraries whose small subset is available for apps, while HALs and thus native servers can use various other vendor-specific native libraries. Linux runs below the Android framework and native libraries (e.g., Android 9 uses Linux kernel v4.4, v4.9, or v4.14). Below the Linux kernel, Android currently supports two different types of instruction set architectures: ARM (32/64 bits) and x86 (32/64 bits).

[4] NDK includes libc.so, libdl.so, liblog.so, and libm.so.

This description contains only a part of how the previous framework was tightly coupled with the hardware dependent code. Specifically, while native servers are part of the framework, they dynamically loaded the system-on-chip-specific HAL code into their address spaces and depended on other native libraries specific to a system-on-chip or device. Framework, native servers, HALs, and native libraries can directly access any kernel interface (e.g., system call and file systems), while the formats of some files in procfs and /dev have differences over time and across various devices. As a result, the previous framework and native server implementations had code logic that can support multiple kernel variants and thus must be verified as a whole using specific kernel and system-on-chip combinations. This poses a technical challenge in decoupling and modularizing the system layer components.

It is also technically non-trivial to enforce system-level interface compliance requirements on finalized Android products that use user builds. That is because Android has used a fully enforced SELinux since version 5.0. SELinux rules restrict arbitrary interactions between the system layer components in user build device. As a result, the Android framework public APIs, which are mostly Java APIs, are mainly available to user apps and non-privileged users. With few exceptions, there are not many known ways to use low layer APIs from user apps.

## 3 PRINCIPLES

Project TREBLE is to actualize the modularized Android platform and make an impact on the Android platform updatability. Our primary research challenge is to derive a modularized architecture design that respects separate component ownerships and thus enables fast platform updates by individual ecosystem entities. Our secondary research challenge is to find a way for an ecosystem entity to develop and propagate all necessary architectural changes to a large part of the Android ecosystem using its yearly release model. These research problems are addressed by these four principles: design, test, coordinate, and deploy.

**Design.** *We modularize the platform stack by repackaging components according to their post-launch maintenance ownerships.*

We first group together components updated by the same entity. Then we keep the ability to independently release each modularized layer by more than one entity including the layer owners who are strongly motivated to make updates. Figure 2(b) shows how the software stack is changed as a result. The redefined layers are caught as system.img, vendor.img, and boot.img. system.img keeps all silicon- and device-agnostic components such as ART, NDK (native development kit), and Vendor NDK (VNDK). VNDK consists of a set of versioned, stable native libraries for vendor interface implementations (e.g., HAL modules). That means if framework vendors or device manufacturers fix any security bugs, then they can update VNDK without having to collaborate closely with vendor.img owning entities. Components in vendor.img (such as HAL and kernel modules) are further modularized. HAL interfaces are reified by using a new hardware interface definition language (HIDL), and then finally redesigned to communicate with the framework across the vendor interface using new IPC mechanisms customized for native code. Also DLKMs are supported so that in addition to silicon vendors, device manufactures can update them as part of a vendor.img update. As a result, boot.img can contain an Android common kernel image but never kernel modules.

**Test.** *We reify the interfaces of modularized components and enforce them by using automated, open source compliance tests.*

To verify the new vendor interface on Android devices, we use our vendor interface test suite, while also using the existing compatibility test suite (CTS) for Android framework interfaces. This concurrent, dual-layer API compliance enforcement allows motivated entities in the ecosystem to use them as integration contracts and then independently develop and update the Android framework to deliver new features, security patches, or bug fixes. Similarly, TREBLE is a basis for other entities (e.g., silicon vendors) to upgrade the vendor partition without breaking the platform or app compatibility. Section 4 further discusses how we identify such API surfaces and cover them as part of automated tests.

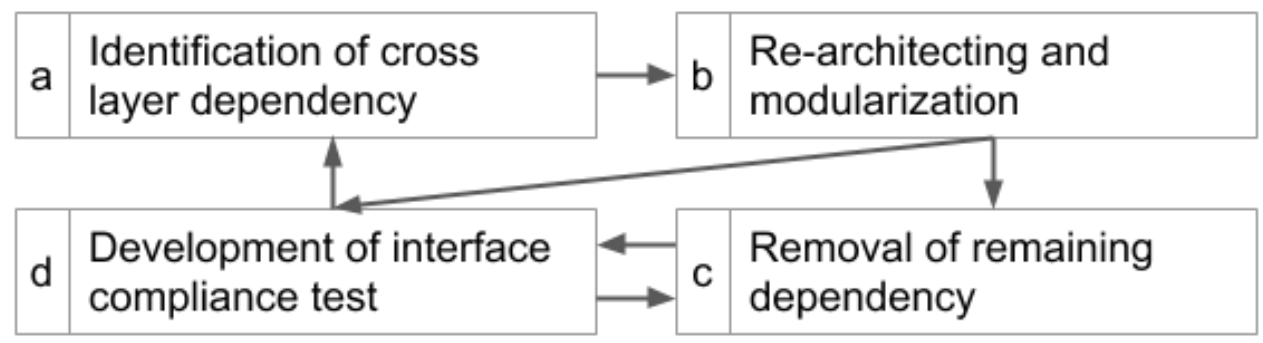

Fig. 3. Process to build a new complete abstract layer.

**Coordinate.** *We heavily invest in development tools to drive all required changes in all component teams simultaneously.*

TREBLE introduces two main types of tools for component and test developments. Component development tools allow each component developer to easily develop modularized clients and servers (e.g., converting a shared library HAL to an independent process with versioned, stable IPC APIs) and integrate with the rest of system. Test development tools are for the same component teams to develop efficiently the vendor interface tests using their domain knowledge. After building the first versions of these development tools, we request and coordinate the required changes across all component teams simultaneously. That coordination strategy not only helps us meet the re-architecting timeline (i.e., HAL conversion in the first year, and completion of other component types by the second year) but also makes it possible for a small engineering team to drive all TREBLE changes in the Android platform.

**Deploy.** *We build a development-deployment pipeline to overlap internal development and external deployment processes.*

We take these outcomes and integrate them with our device partners in the ecosystem as part of new Android platform code.[5] In practice, the internal development and partner deployments have a large time overlap. That allows Android device partners to have an enough room to integrate our architectural changes and ship them to users within the same yearly release cycle.

## 4 DESIGN

TREBLE architecture aims to modularize Android in such a way as to allow for separate ownership of parts of the Android distribution according to framework and silicon vendors and device manufacturers. A major part of this modularization is the reifications of the HAL, vendor native library, and the Linux kernel interfaces to the status of versioned and standardized APIs to the framework, as well as the creation of comprehensive system-level tests to ensure compliance of these new APIs.

### 4.1 Interface Reification

As depicted in Figure 3(a), we identify how the Android framework can directly interact with its underlying vendor interface implementation partition. We then reify all identified interfaces to form the vendor interface:

[5] Called dessert release because versions are named after desserts.

***(i) HAL.*** HAL interface wraps various I/O device drivers. Those HALs take a large portion of the vendor interface surface. We *binderize*[6] those HALs with a few exceptions. Specifically, we modularize HALs as independent processes running in a vendor partition, formally define their APIs, and make the framework use an IPC mechanism to communicate with them as shown in Figure 2(b).

Similar to the existing AIDL (Android Interface Definition Language) for Android framework layer IPCs, we use our new *Hardware Interface Definition Language (HIDL)* to specify HAL IPC interfaces. Our HIDL is suitable for specifying native layer communications (e.g., shared memory between C/C++ clients and servers). An interface in our HIDL has a set of remote procedure call (RPC) APIs where each API has a set of typed arguments and return values. Typically, a HAL is implemented as a Linux service, and a native server becomes the client of each HAL server. HIDL supports user-defined data types that use basic data types commonly used in native code and provide intuitive memory and control flow models to its developers. HIDL has a versioning scheme and ensures the ABI stability required for the decoupling. A HIDL client-server pair typically uses *hardware binder (HwBinder)*, which is our new binder implementation customized for HIDL data types.[7]

A *binderized HAL* no longer lives in the address space of its client, e.g., a native server (see Figure 2(a) and 2(b), the native server address space blocks). If a HAL was attached to a native server, it gains an additional RPC call from the native server to its new HIDL HAL server. Alternatively, if a HAL was directly attached to the Java framework through JNI, the Android framework can directly call the new HIDL HAL server APIs via JNI using a library that knows how to convert a Java function call to an HwBinder call. Some HALs are converted to use HIDL but not used as binderized HALs (namely, *pass-through HALs*) because direct function calls are still required due to the system-wide overheads that are unavoidable if those pass-through HALs (e.g., graphics-mapper and renderscript) are configured to run in binderized mode.

While the vendor interface for HALs is reified using HIDL and HwBinder, the existing non-HAL interfaces fall into two categories: stable non-HAL interfaces that provide the same properties as HALs, and unstable non-HAL interfaces. Stable non-HAL interfaces include GL/EGL and OpenMAX and provide stable, named, binary, versioned interfaces; thus some of them are kept without requiring any HIDL conversions. Unstable non-HAL interfaces are *legacy HALs* that are typically a static or shared library and do not define any of the symbols that the framework requires for API lookup. Legacy HALs are telephony radio interface layer (RIL), Wi-Fi, and Bluetooth interfaces to the hardware. All legacy connectivity HALs are converted to binderized HIDL HALs as a result of TREBLE.

Unlike the Android public API level the Android framework provides to apps (e.g., API level 28 for Android 9), the vendor interface version is not encapsulated in a single number. Instead, the vendor interface version is a concatenation of the versions of its components (e.g., HIDL HALs). The *vendor interface object* is a read-only entity that is queryable from both native code and from Java, which can report the entire version string.

***(ii) Vendor Native Library.*** Vendor native libraries that are not part of NDK (native development kit) were used by both the Android framework (e.g., native servers) and vendor interface implementations. Those vendor libraries were logically part of both Android system and vendor partitions. Thus, such libraries tightly coupled these two partitions as depicted in Figure 2(a). To decouple them, we present the *Vendor Native Development Kit (VNDK)* approach.

We formally define VNDK as a set of native libraries commonly used by HALs. VNDK is kept in the system partition (as shown in Figure 2(b)) but is mainly for vendor partition use cases with a few exceptions for pass-through HALs on the system partition. As a result, upgrading system.img also upgrades all the VNDK pieces, making silicon vendors free from such maintenance burdens.

[6] verb. Convert software to use a binder-like IPC mechanism.
[7] We omit individual techniques in this paper to focus on the major architectural contributions made for fast software updates.

VNDK has versioned, stable APIs. The VNDK version has two components: major and minor versions. The major version is the same as the version of the framework that has the released VNDK. The minor version is the security patch level. We maintain ABI stability within each major VNDK version. Within each major version, we update the VNDK in a binary-compatible manner. We call a stable, ABI-compatible instance of the VNDK *a VNDK snapshot*. We maintain a set of VNDK snapshots, one per major framework version (e.g., v10), and deprecate them on a sliding-window basis. For example, in Android 10, we support VNDK v8.0, v8.1, v9, and v10. When Android 11 is released, support for VNDK v8.x is dropped, and vendor implementations still get stuck at v8.x cannot survive without a VNDK snapshot v8.x that is no longer supported.

VNDK lives in new subdirectories (i.e., /system/lib*/vndk-<API level>) in a system partition. That makes it possible to keep only one copy of vendor native libraries for vendor and system partitions because code in a vendor partition is also allowed to depend on them. A VNDK distribution is conceptually similar to an NDK distribution. For every Android framework version, VNDK contains prebuilt binaries with symbols of all the libraries it tracks, one per instruction set architecture. The VNDK snapshots differ for each major architecture based on its hardware support options. There are over twenty variants across the four major architectures. The actual binaries shipped depend on what the hardware reports through the vendor interface object.

***(iii) Kernel.*** We define the Android *common kernel interface* and further modularize the kernel layer software.

*(a) Common kernel interface.* The kernel portion of the vendor interface consists of core and Android kernel interfaces:

(1) Core kernel interface. Android already had specific requirements for the Linux kernel. Requirements range from kernel features that are part of the main Linux kernel distribution (such as cgroups) to various standard Android additions to the baseline kernel, whether or not they have been upstreamed (e.g., binder, ION, and sync framework). Standard Linux kernel functionality is naturally exposed in native code and in Java (e.g., cgroups exposed to android.util.Process via /sys/fs/cgroup).

(2) Android kernel interfaces. Another set of kernel interfaces implement system-on-chip- or device-specific functionality. Access to such custom interfaces is already well separated into libraries, regardless of whether they conform to the hardware abstraction layer interface. There are a few exceptions to this rule, such as in the implementation of android.os.Debug.getMemoryInfo, which has direct references to system-on-chip-specific interfaces in /dev). Issues such as getMemoryInfo are resolved by migrating the code into an existing or a new hardware abstraction layer module. For interfaces that are accessed either directly or with utility libraries, we either remove the utility library from use or add it as part of new low-level NDK (LL-NDK) accessible to both system and vendor code.

*(b) Kernel modularization.* We use three major techniques for modular kernel layer software. (1) DLKM. A large part of kernel security and reliability defects was found in device drivers [3]. We use DLKM and contain those device drivers in vendor.img that silicon vendors have the ownership of maintaining and updating that partition. (2) Device tree overlay (DTO). We also require device tree supports in Android kernel. This allows various entities to describe system-on-chip- and board support package (BSP) specific devices in the form of a device tree overlay. Device tree overlays are also not part of boot.img but kept as part of a new dedicated partition for independent updates. (3) SELinux policy split. We keep two sets of SELinux policy files: one in vendor.img and the other in system.img. At boot up, they are merged into one set and then used to enforce the unified access control policies. As a result, it is possible to update SELinux policies that are needed to accommodate any changes made in new Android frameworks just by changing and updating system.img.

***(iv) Extended API.*** The TREBLE architecture introduces a new vendor extension partition, namely, original design manufacturer (ODM) partition. Using this extension partition, Android device ecosystem entities can extend default HAL interfaces, create new HAL types, and ship them as part of an ODM partition. *(a) HAL customization.* Let us assume Best Manufacturer wants to extend an interface of android.hardware.light@1.0 HAL package. Then Best Manufacturer defines an interface in vendor.bestmfr.light@1.0 where the defined interface is inherited from one in android.hardware.light@1.0 and has new data types and APIs that are specific to the light hardware device of Best Manufacturer. *(b) New HAL type.* If Best Manufacturer supports a hardware device that is not available in AOSP, then it defines a new HAL package (e.g., vendor.bestmfr.microprojector@1.0) and can maintain multiple interfaces and versions in it.

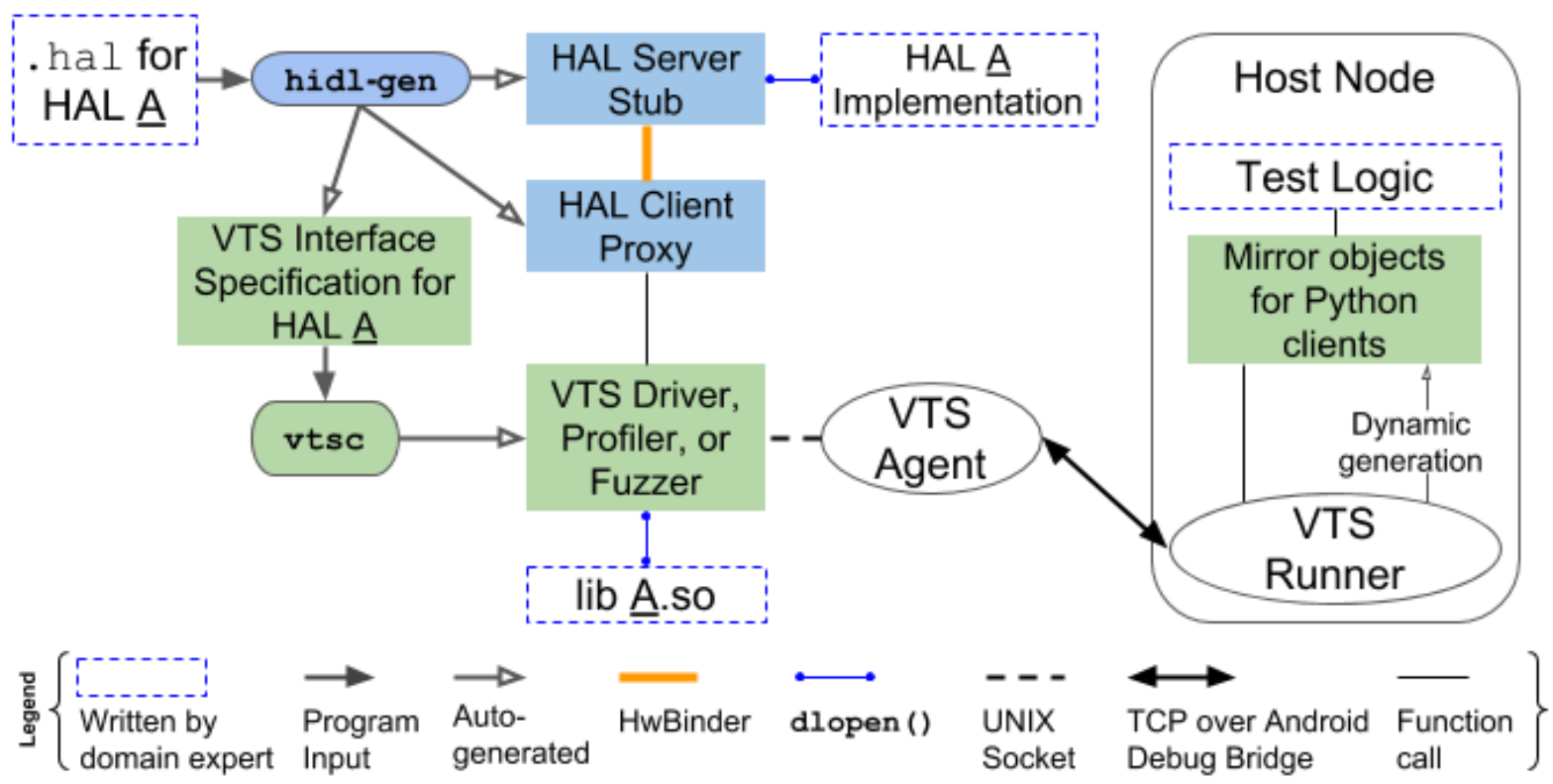

Fig. 4. Auto-generation of HAL middleware and test components.

## 4.2 Interface Verification

As shown in the last step of Figure 3, we develop and use a set of interface tests to ensure that a given vendor interface implementation is fully decoupled from the Android framework and only interacts with the framework using the three primary vendor interface surfaces that we define and modularize in Section 4.1. Specifically, that goal is realized by our two new compliance test requirements. All new vendor interface implementation must pass the following: (1) vendor interface test suite (VTS requirement), and (2) existing compatibility test suite while running an Android generic system image (CTS-on-GSI requirement). To this end, vendor interface test suite is designed to include test cases for all the components covered by the vendor interface (i.e., HAL, VNDK, and kernel). The CTS-on-GSI requirement further *ensures the backward compatibility of pure Android frameworks*. Those two rules are the basis for guaranteeing that a tested vendor interface implementation can execute both current and future Android frameworks and still pass compatibility test suite, an indicator of app compatibility in the Android application ecosystem.

***(i) HAL.*** Our test framework is specifically designed to support both target-side and host-driven hardware abstraction layer module tests.

*(a) Target-side tests.* A target-side HAL module test uses a device-side test binary, pushes that binary to a target Android device, and executes that binary on the device. The host-side test framework stops the Android framework and relevant native services while running its test cases because that helps us find any unexpected framework and vendor dependencies and HAL modules are typically not designed to serve more than one client at a time.

*(b) Host-driven tests.* Figure 4 shows the compilation and test execution flows of a HAL test. A developer writes a HAL interface specification file (.hal file). Then the HAL compiler (hidl-gen) automatically generates HAL client proxy and server stub code. hidl-gen also generates a .vts intermediate file that is used by the compiler of vendor interface test suite (vtsc), which in turn automatically generates a driver, profiler, and fuzzer of a given HAL interface. A test driver dynamically loads a HAL client proxy shared library and calls the APIs of the loaded library when a request sent from a host-side test case arrives through the test agent running on a target Android device (see Figure 4). Using this remote procedure call (RPC) mechanism, test writers can develop host-driven HAL tests where the main test case logic is kept and running on the host-side. The required host-side RPC mirror objects are dynamically generated at test runtime from a given .vts intermediate interface specification file. That dynamic generation is done by using Python reflection because the host-side test runner (see Figure 4) is written in Python. In general, host-driven tests are recommended when test writers want to mix various test scenarios (e.g., native and framework layer operations or multi-device testing) in the same test.

HAL profilers are useful to measure the binder-ization overhead of HAL modules. A profiler can transparently measure the performance and record all API calls of a HAL while running any of its existing HAL tests. This is possible because our test framework at test runtime decides how to access a target HAL service (e.g., using HwBinder or pass-through mode) and thus can add instrumentations. Using the same mechanism, we collect HAL API call traces while running compatibility test suite against an Android framework. The collected traces become test cases after certain post-processing, and also help us select a subset of test cases for the CTS-on-GSI requirement (e.g., by measuring HAL API coverage metrics). That subset is used for CTS-on-GSI because running compatibility test suite can take more than three days (without test sharding on multiple devices) and reducing the compliance testing time is important for the productivity of Android device partners.

Basically, any CTS module that does not exercise any HALs during its execution can be safely removed from the CTS-on-GSI requirement. We use HAL profiling to measure what kinds of HAL APIs are called. It alone however does not tell whether the initiating entity of each HAL API call is a target CTS module. We thus collect HAL API call traces from an idle device and then use them as background noise data. For example, let us assume that profiling of a CTS module generates a HAL trace file for HAL A. If the background noise data also includes the trace of HAL A and the trace file size is on the same order of magnitude, then that CTS module does not intensively exercise HAL A and can be removed. Because HAL API call traces are non-deterministic, this procedure provides us a list of CTS modules that have relatively long testing time (i.e., to maximize the benefit of removal) and a relatively large overlap with the background noise data (i.e., to minimize the testing coverage lose). After the data is provided, the final decision is left on domain experts (e.g., each CTS module owner).

Similarly, HAL fuzzers are critical to enhance the robustness of a given HAL module implementation and consequently reduce the need for updating HAL modules in the vendor partition. Our HAL fuzzers are auto-generated and use libFuzzer, ASan, and SanCov greybox fuzzing tools to test given HAL interfaces in an effective way.

***(ii) VNDK.*** Vendor NDK interface verification is done in three ways: *(a) Runtime checking.* Linker namespace is used to prevent dlopen() calls against non-VNDK libraries from vendor partitions to a system partition. In the implementation, it uses the existing dynamic loader extension made in Android 7. That extension loads the ld.config.txt file to find the specified linker namespace rules and create namespaces accordingly. *(b) Vendor interface testing.* Tests written for VNDK provide extra vendor interface coverage. A test checks whether a VNDK snapshot depends only on NDK and other VNDK libraries by pulling every VNDK library file then extracting and analyzing its file dependencies encoded in Executable and Linkable Format (ELF). *(c) Static application binary interface (ABI) checking.* For a system partition only OTA (over-the-air) update, one needs to be able to check the ABI compliance between the vendor interface implementation of a target device and the new VNDK being sent. There are three known ways to check ABI compliance: extracting from source code, headers, and binaries. The source code-based technique is the most accurate because the debugging information in a binary is stripped to reduce the binary size. Our source code technique is integrated with the build system for continuous enforcements during development. Another tool also collects the ABI information of headers and compare with that of respective VNDK snapshots (e.g., one equivalent to an old VNDK version on the target device). Lastly, binary checking is done as part of compliance tests (i.e., the VTS requirement).

***(iii) Kernel.*** Vendor interface test suite is used to test the kernel interface. To this end, we extend our VTS test framework to support Android-specific kernel testing. Most of these are written as host-side tests in Python. We also port two common Linux kernel test suites to Android and the ARM instruction set architecture: Linux Test Project (LTP) and linux-kselftest. Both are used widely in the upstream Linux community. These two tests are useful for ensuring the POSIX compatibility.

***(iv) Extended API.*** It is recommended for Android device ecosystem entities to write VTS tests for HAL extensions and vendor-specific HALs they create. Similar to the VTS tests for the default HAL interfaces, those extended tests can ensure the API stability of an extended HAL interface surface. Those extended tests are also useful if an ecosystem entity extends the AOSP framework in a way that the framework extension directly uses the extended HAL API surface. That is because the extended tests can help ensure the backward API compatibility of the extended framework. The ensured backward compatibility property makes it easy to take new vendor and ODM images as part of an Android platform version upgrade.

## 5 EVALUATIONS

We use some state-of-the-art Android smartphones for prototyping and apply our techniques to the Android device ecosystem. We have tracked the update velocity of multiple Android device makers from 2016 to 2019. Our experiments and industry rollouts confirm that the TREBLE architecture is practical and is a strong foundation for enabling the fast Android platform software updates.

### 5.1 Updatability

In this experiment, we prepare and use three Android phone models. The devices use three different system-on-chips manufactured by two different silicon vendors. All three phones use their production images (i.e., user build images), which are not rooted or modified. We build a generic system image using aosp_(arm|x86)[64][_a[b]] Android lunch targets that we create in the Android Open Source Project where '_a[b]' means a seamless update using two images, A and B. We flash that same Android system.img to the three different kinds of Android phones. In early 2017, our experiment first showed all the phones booting properly and meeting our compliance requirements without any notable reliability issues. Since our initial experiment, other entities in the Android ecosystem have repeated this using various smartphone models and reported the same. The fact that we can use the same pure Android framework image on three different vendor implementations confirms the modularization result (i.e., split of core OS and the silicon- and device-specific code). This experiment further validates that our methodology of finding the cross-layer dependencies, decoupling them, and enforcing the API compliance layers is feasible and effective because we accomplished it within one Android platform release cycle (Android 8.0 and 8.1 releases in 2017).

The presented TREBLE architecture is released as part of the Android Open Source Project since version 8.0. Since then, the TREBLE architecture has been successfully adopted by all new Android 8.0, 8.1, and 9 devices with Google Mobile Services [21]. In 2019, Android 10 continues to use the TREBLE architecture [22]. Specifically, Android 8.0, released in mid-2017, is the first Android platform version that partially uses TREBLE. Between mid-2017 and early 2018, the following flagship Android devices are successfully launched with TREBLE: Samsung Galaxy S9, Google Pixel I and II, Huawei Mate 10, Sony Xperia, and dozens of Android Go phones [23]. The Android 9 release a year later fully uses TREBLE. Thanks to the basis created in Android 8.x, in addition to

Table 2. User Perceivable Camera Quality Metrics

| Video resolution (fps) | Camera frame drop rate | | Video resolution (fps) | Video recording charge (13 minutes) | |
|---|---|---|---|---|---|
| | Base | TREBLE | | Base | TREBLE |
| **HD 1080 (30)** | 0.02% | 0.01% | **Preview (≥15)** | 88.8 mAh | 87.4 mAh |
| **HD 1080 (60)** | <0.01% | <0.01% | **HD 1080 (30)** | 143.8 mAh | 149.6 mAh |
| **UHD 4K (30)** | 0.04% | 0.02% | **HD 1080 (60)** | 197.8 mAh | 206.0 mAh |

Google, on May 8, 2018, seven Android device manufacturers (Essential, Nokia, OnePlus, Oppo, Sony, Vivo, and Xiaomi) have demoed Android 9 previews on their previous generation phones, which were initially launched with Android 8.0 or 8.1 [24]. Because the first developer preview release of Android 9 was on March 7, 2018, this clearly demonstrates that fast updates of existing smartphones to a new Android platform release are possible. In 2017 when TREBLE was not available in previous generation phones (e.g., Android 7.x), no device manufacturer other than Google Pixel demonstrated the same (i.e., fast Android framework update).

The benefits on mobile software users are clear. Android device partners can directly send their software updates to mobile users for bug and vulnerability fixes, and to enable new features and services on legacy hardware devices. The modularized mobile platform architecture presented in this study allows every entity in the Android device ecosystem (e.g., device manufacturers, silicon vendors, and framework vendors) to initiate such mobile software updates independent of the others as long as their users are willing to take the provided mobile software updates.

### 5.2 Performance and Power Overheads

Historically open source software communities are in favor of component-based design because of a large number of distributed contributors who can easily face the high costs of communication and integration. However, the associated performance overhead limited [6] extensive adoption of this design in certain embedded systems. Our TREBLE architecture thus relies on the efficient IPC primitives specifically designed for Android native components. Our evaluation using Google Pixel phones with a quad-core CPU (two cores running at 2.15 GHz) and 4 GB RAM shows that these primitives well optimize the user perceived performance and power overheads. The measurements are done by using the performance and power tests of VTS (i.e., an optional part).

***(i) Application Metrics.*** We measure two main smartphone quality-of-service metrics:

*(a) Camera.* The camera frame drop rate is measured while recording three kinds of videos. We do not notice any major regression in any of the recording periods where each period lasted for five minutes (see Table 2). We used five minutes because the default camera app of Android 7.0 used as the baseline has a low limit in the video recording file size and five minutes cover most of the typical use cases. Because the camera frame drop rate metric is highly sensitive to various conditions (e.g., temperature and actual camera scene), our tests include steps to cool down devices and stop measurements if their preconditions are not met. With other optimizations done as part of Android 8.0, Pixel phones with Android 8.0 did not regress compared to the same phones with Android 7.1. That also required some TREBLE-specific optimizations. For example, the frame drop rate was notably improved when we properly used the Linux FIFO scheduler with the binder-ized camera and media HALs. Because a 0.04% frame drop rate is generally considered an acceptable release condition, no further optimization was required in the prototyping stage. The following power measurement experiment uses thirteen minutes to analyze the impact of longer camera usage scenarios.

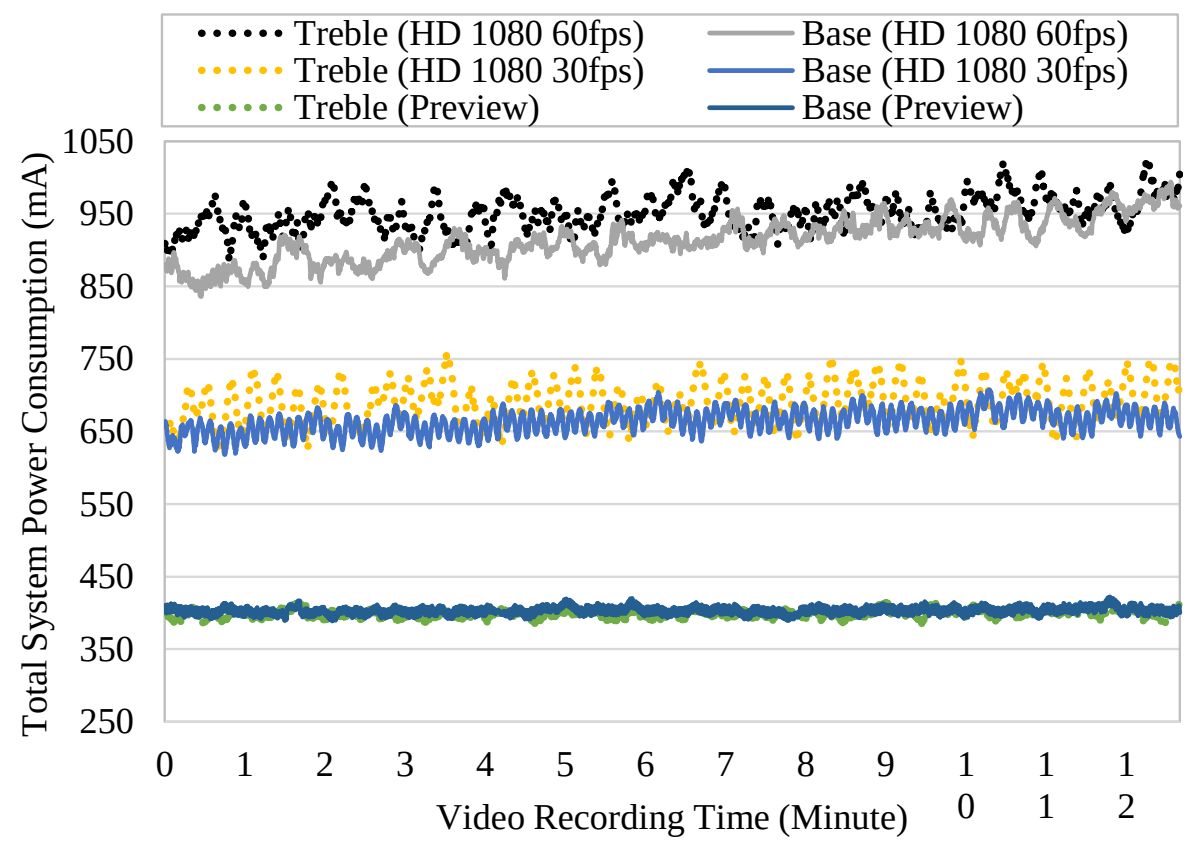

Fig. 5. Video recording power consumption vs. recording time.

In terms of power consumption, we saw an increase in the power consumption mainly due to the extra inter process communications. Specifically, we saw 2.2% increase in power consumption during the HD (high definition) 1080p x 60*fps* video recording tests (908mA vs. 928mA). We saw slight improvements in the other cases (i.e., photo preview and HD 1080p x 30*fps* recording) during the prototyping. After the first devices using TREBLE were launched, we took the over-the-air images and measured the total charges while recording for thirteen minutes. As shown in Table 2, the photo preview operation reduces the total charge consumption by ~1.6%, while HD 1080p recordings increase the total charge consumption by ~4% for when *fps* is 30 and 60. Less than 5% overhead under some specific use scenarios was expected given the updatability and quality benefits. We note that this power measurement experiment is conducted indoor where ambient light is controlled to minimize the impact of a camera auto exposure algorithm.

Figure 5 shows the total system power consumption as a function of the video recording time. It uses the over-the-air update images of Android 7.1.1 and 8.1 and omits the data for first ten seconds. We observed relatively larger variations with a user build Android device running TREBLE-ized Android 8.1 than the same device running a non-TREBLE-ized Android 7.1.1. In a heavy task (HD 1080p x 60*fps*), the stabilization point was beyond ten minutes (vs. about six minutes in case of HD 1080p x 30*fps*). That is because of the device temperature increase which negatively impacts on the power consumption. In that case, both devices suffered from the thermal issue and the average power consumption of both devices becomes similar after eleven minutes of video recording.

*(b) Audio.* We conducted an audio loopback latency test where a rig [25] was used to connect the audio speaker output sound to the microphone input jack. Each measurement is done in less than a few seconds and the experiment is repeated for twenty to thirty times. Our baseline Android device shows a 16.0 ms average latency where the standard deviations are 3.4 ms. Our prototype using the same Pixel phone model but using new Android platform software (pre-8.0) with our techniques and other changes showed 15% increase in the average latency and a ~1.3 ms increase in the standard deviations. It showed that the jitters are more significant as a result of the modularization and binderization. Our prototype used a shared memory-based message queue to improve the audio latency further. The final optimized solution has more than 90% of the latency samples within the 20 ms threshold which is known as user perceptible. After that prototyping stage, further improvements were made that helped various device manufacturers listed in Section 5.1 pass QA and ship their Android 8.0, 8.1, and 9 phones.

Table 3. HwBinder Roundtrip Latency

| Message | Best Case | Average | Std. Dev. |
|---|---|---|---|
| **4 B in size** | 65.7 µs | 66.7 µs | 0.8 µs |
| **2 KB** | 67.7 µs | 69.3 µs | 1.1 µs |
| **4 KB** | 79.4 µs | 80.7 µs | 0.7 µs |
| **16 KB** | 107 µs | 111 µs | 3.7 µs |
| **64 KB** | 270 µs | 277 µs | 4.2 µs |

Table 4. HwBinder Throughput

| Concurrency | Average | 90 Percentile |
|---|---|---|
| **2 pairs** | 98.2 µs | 195.3 µs |
| **3 pairs** | 162.3 µs | 215.9 µs |
| **5 pairs** | 215.4 µs | 579.1 µs |
| **10 pairs** | 370.8 µs | 825.8 µs |
| **100 pairs** | 3.9 ms | 13.3 ms |

***(ii) Micro-benchmarking.*** We use our microbenchmarks to characterize the performance of our IPC primitives.

*(a) HwBinder performance.* Table 3 shows the roundtrip latency of HwBinder as a function of the message size. The raw performance for a 4 byte message (e.g., equivalent to a function with an integer argument value) is 66.7 µs on average. The average latency is stable if the message size is smaller than 4KB, a memory page size. As the message size becomes larger, the latency becomes naturally longer due to the memory copy overheads.

*(b) HwBinder throughput.* We measure the throughput of HwBinder when multiple clients and servers simultaneously communicate using the HwBinder kernel driver. Table 4 shows the average and 90 percentile of roundtrip HwBinder call latencies when multiple client-server pairs simultaneously exchange 512 byte messages. It shows the average latency grows reasonably until around 10 client-server pairs thanks to the various latest kernel-level optimizations (e.g., uses of fine-grained locks instead of the global lock in the HwBinder kernel module).

*(c) Shared memory solution.* The average latency to read a 64 byte and 512 byte message using a shared memory API is only 217 ns and 274 ns, respectively, on a 32-bit native application. The read and write performance was almost symmetric although 32-bit write performance was 8% faster than 32-bit read performance when the message size is 512 bytes. The shared memory IPC is the best option in Android native layer if a large size memory copy is needed.

### 5.3 Impact on Quality and Fault Localization

We have continuously run the tests in vendor interface test suite for over two and a half year and found several hundred bugs, some of which are critical vulnerabilities. Our data shows that vendor interface test suite finds many critical bugs in the vendor partition and thus reduces the need for post-launch updates of the vendor partition. The defect statistics confirm that directly testing the low layer APIs finds more problems than solely relying on upper layer app UI testing. Because part of our tests (e.g., kernel tests) were run at presubmit time and our partners who also use these tests did not report all the bugs they found in their own Android extensions, we expect that the actual count of bugs found or prevented by our interface tests is larger than what we have tracked.

Table 5. New Git Projects of TREBLE

| Type | Git Project Path |
|---|---|
| **HAL** | hardware/interfaces, system/tools/hidl, system/[libhidl, libhwbinder, libfmq], */hardware/interfaces |
| **VNDK** | prebuilts/vndk/* and embedded in various existing AOSP git projects |
| **Kernel** | kernel/configs and embedded in Android common kernel repository |
| **VTS** | test/vts, test/vts-testcase/*, system/libvintf, external/ltp, external/linux-kselftest, test/framework, test/vti/*, tools/acloud |

In most cases, when a vendor interface test case fails, it correctly identifies the location of a problem. It not only tells us that a bug is below the vendor interface layer but also identifies the specific module with the defect. In cases where all vendor interface test cases of a module pass, it typically localizes the software faults. For example, upper layer tests (e.g., CTS) failed because the problem was likely above the vendor interface and not in a component of the vendor interface implementation that was relatively well covered by the used vendor interface tests.

## 6 DISCUSSIONS

This section discusses other development, integration, and deployment aspects of TREBLE and summarizes the key lessons we learned and the implications for other embedded software platforms.

### 6.1 Software Engineering Cost

We describe how we deal with the challenges associated with the scale of our development and integration tasks. In total, ≥25 new git[8] projects are created in Android Open Source Project where ≥25k commits and ≥2.2M lines of changes are made in the TREBLE first year (for Android 8.0 and 8.1) and ≥27k commits and ≥2.6M lines of changes are made in the second year (for Android 9). These statistics are collected conservatively. Table 5 provides a summary of the key project list.

***(i) HAL.*** Our team helps component development teams to perform HAL conversion and vendor interface test developments (see Figure 6, the flow from initiator team to component teams). In total, about 35 HALs are successfully converted in the first year that include legacy non-HALs (e.g., telephony). Vendor interface tests are written for over 50 HAL versions. While most HAL tests cover all APIs, only a few HAL tests have achieved above 70% HAL code line coverage. In total, over 70 different engineers worked on the HAL conversion and test developments.

A HAL conversion not only requires its developers to specify all callable APIs of a HAL[9] but also callback APIs implemented on the client side. As a result of conversion, some HALs have significantly increased API counts. For example, a HAL that uses a pipe-like API to pass various commands to its kernel driver needed to be rewritten in a way that explicitly specifies an API for each command type. In some cases, when .hal files were written, both developers and testers started developing in parallel, especially when more than one engineer was assigned to a HAL and their responsibilities were clearly defined.

[8] Git is a distributed version control system and widely used by Android, Linux, and various other free open source software projects.

[9] Needs the definition of a list of methods per interface. A method has a name, typed arguments, typed return values, and annotations..

***(ii) VNDK.*** We handle the libraries we own and third-party libraries differently. Our analysis of common libraries identified 164 total native libraries, where 45 were third-party libraries and 119 were internal libraries. In terms of code location, 62 libraries were under frameworks/ and 57 were not, while 32 were under system/ and 13 were under hardware/. Because many libraries under frameworks/ are internal to the Android framework, we exclude them in our VNDK work. In practice, an upper limit of the size of the VNDK is the size of /system/lib* on the build server, which was about 200MB for a double-architecture userdebug build target.

***(iii) Kernel.*** We first focus on ensuring big structures such as the kernel config options (using /proc/config.gz) and a list of required proc files. Most of our work on the kernel vendor interface was in identifying the direct dependencies between the framework and kernel. We identify them by using dynamic and static techniques. *(a) A dynamic technique* uses a set of modified SELinux rules and audits all accesses to various kernel file systems (e.g., procfs, sysfs, /dev, selinuxfs), while running compatibility test suite and top Android Play store apps. The system calls used by the Android framework are identified by enabling seccomp for a certain period. The seccomp Linux kernel feature allows a user process to switch to a secure state where it can perform only a limited number of syscalls. *(b) A static technique* relies on searching the source code using specific terms (e.g., /proc). We then communicate with the owners and authors of the code that uses the identified interfaces because it is a fast way to use domain expertise required for the task. Domain owners write vendor interface tests for the content of each identified critical file, typically using regular expressions. If a kernel exposed file is not supposed to be accessed by the Android framework, then its access is constrained by SELinux rules.

***(iv) VTS Tests.*** Vendor interface test suite for Android 8.0 has 2,000+ HAL test cases, 1,000+ kernel test cases, and 2,000+ library test cases. While most test cases were manually written by domain experts inside and outside of our organization, the rest were automatically generated by using record-and-replay and fuzzing techniques. Vendor interface test suite has an optional part (namely, VTS-*) that includes nonfunctional tests, many of which were also automatically generated and have been critical in finding defects alongside HAL, kernel, and library structural tests.

### 6.2 Deploying to Device Partners in Ecosystem

We discuss how we deploy the developed architectural changes to our device partners in the Android device ecosystem (see Figure 6, the outcome of initiator organization) and collect feedbacks.

In mid-2016, after we finished our design and were doing our tool prototyping, some team members visited key partners to share our visions and collect their opinions. Partners who wanted to ship new Android devices using our architectural changes and rewritten components needed to rewrite their components (e.g., extensions) and do integration works. As partners started to use the provided components and tools, they sent feedback that our teams continuously addressed. We had shared repositories and an issue tracking service for partners to track reported issues. This helped us resolve various complex engineering problems that are often difficult to solve without seeing the actual artifacts.

One technical challenge is that vendor interface tests must be run against all relevant final products produced by partners. While final products are user build devices (i.e., non-rootable), running system-level vendor interface tests requires an extra privilege (e.g., HwBinder call requires an SELinux rule change). We provide a *reference Android framework image* (i.e., generic system.img) and ask partners to flash their final products using that provided image. In that way, we are able to run system-level tests against the user builds vendor interface implementations of partners.

### 6.3 Lessons Learned

The key lessons we learned during the TREBLE transition are:

***(i) Component Rewriting Coordination.*** We share best practices across the component development teams by building and taking advantage of a horizontal organization. As a medium size integration team, we interact with many development teams with large numbers of engineers (e.g., several hundred engineers in over thirty teams). Thus, development coordination efficiency is important to keep the size of our integration team and compress the product launch schedule. Basically, *nominating an engineer* from our integration team to *each domain* (e.g., media, graphics, connectivity, security, and IO) is effective because then the nominated engineers learn domain expertise and build rapports with the respective component teams. The nominated engineers form a *horizontal organization* that holds frequent meetings to *help* them *identify best practices and common mistakes* and rapidly propagate to other domains (e.g., by enhancing tools or code review). It also helps them identify the most critical features of the development tools and infrastructure they provide. These tools and infrastructure developments are done incrementally using an agile model that helps us *compress the project schedule by overlapping the tools and infrastructure developments, and the actual component rewriting and integration*.

After the baseline infrastructure and processes are in place, it is important that we *change our coordination model to avoid such direct human engagements whenever possible* and practice other low cost coordination approaches for various tasks. The best case is when our tools prevent anticipated problems and naturally guide component teams to follow recommended design patterns (e.g., as part of automated code review or self-checking). Another case is when one engineer can handle the same issues across entire domains without requesting anything from the component teams other than getting code reviews. Lastly, documentation is another way to scale and sustain non-automatable processes. We produce *one set of documents for both internal developers and external device partners*. That is possible because all our tools and infrastructure code is open source and shared immediately with all external entities.

***(ii) Partner Engagement and Deployment.*** It was crucial to help major entities in our ecosystem get, use, and extend the major artifacts of TREBLE in one Android release cycle (i.e., the first year). *Transparency and openness to feedback* (e.g., code and document sharing, face-to-face meetings, and running a partner issue-tracking service) is fundamental in keeping mutual interests and trust with other ecosystem entities. Such *close interactions* help us discover the interests of some partners in TREBLE due to other technical challenges they face. For example, major device manufacturers use more than one system-on-chip for the same flagship phones (e.g., MediaTek, Qualcomm, and Samsung System LSI chipsets). TREBLE modularization helps these manufacturers reduce their software engineering costs because new vendor interface implementations they receive from the major silicon vendors follow the same API compliance requirement (e.g., VTS).

***(iii) Implications for Other Platforms.*** Other mobile and embedded platforms can take advantage of our TREBLE software architecture design and artifacts. One can reuse our vendor interface compliance tests to split hardware-dependent and independent code as long as their platforms are Linux-based. Using our vendor interface testing tools and infrastructure can help further reduce their test development and other associated software engineering costs. One can also reuse our HIDL and HwBinder IPC middleware to help develop versioned, stable APIs that are backward compatible. The concept of VNDK is also new and can be applied to other platforms if the entities that maintain layers equivalent to Android system and vendor partitions are different. Various kernel modularization techniques we use are good examples for other Linux-based embedded systems that require separate ownership and fast software updates.

We hope that the underlying tradeoffs touched by the TREBLE principles and techniques are well considered not as an afterthought but from the beginning of a new embedded software ecosystem. Advances in the systems technologies enable us to use those TREBLE techniques in various embedded systems as evidenced by TREBLE, which is also used by low-end Android devices, such as Android Go phones with 512MB RAM.

## 7 RELATED WORK

Many recent works studied the update-related security issues of Android apps focusing on app lifecycle [7], third-party native libraries contained in apps [8], and application-layer API compatibility issues [9][10]. By nature (e.g., involved entities), those are different from the platform updatability challenge.

At the platform level, dynamic OS update techniques [11][12] and Android-specific hot patching [13] were studied. That hot patching technique is only applicable to the user space platform code and requires the root privilege. Other works (e.g., virtualization [14][15], containers [16][17], and OS personality [18]) hosted multiple OSes on the same device to provide a larger set of platform APIs and host various kinds of applications. Most of those works focused on API compatibility between different OS types, while we focus on maintaining compatibility within the same OS type.

Another approach of improving the platform update velocity is investing in and using software merging techniques [19], which can partially automate downstream merging processes. A data mining technique [20] was used to improve the source code merge conflict resolution process between Android and its downstream platform. Those are orthogonal to our system software architecture approach.

In terms of actual software techniques, the novelty of our TREBLE techniques is mostly from the unique requirements of the Android platform and its ecosystem model. For example, the necessity of stabilizing and versioning native libraries was shown by the dynamic-link library (DLL) of Windows OS. TREBLE takes this a step further. Because one of our goals is to enable two different types of entities to distribute vendor and system partitions independently, our VNDK is hosted in the system partition and used by code in the vendor partition. While doing so, the existing mandatory access control mechanisms (e.g., SELinux) of Android enforces extra restrictions in designing and testing our system (e.g., resolved by build and test time checks that require our generic system image). In terms of IPC middleware for HALs, we customize the binder for native use cases and enforce versioned, stable HAL APIs using Hardware IDL (Interface Definition Language) because the original Android IDL was neither stable nor versioned [4]. We then enforce the backward compatibility of those reified vendor interface APIs because that enables independent framework only updates of Android platform.

## 8 CONCLUSION

In the past, creating a new software ecosystem was in many cases seen as an easier strategy than maintaining and transforming a large established ecosystem. That was because a key factor that supported the success and fast growth of a software ecosystem can later become a factor that makes it hard to maintain the ecosystem.  Thus, after a software ecosystem reaches a certain scale, its organizational model needs to be adjusted or changed so that maintenance can remain cost effective and the ecosystem can continue to grow and fulfill its mission. This paper provided evidence of our claim in the world largest computing device ecosystem by demonstrating how such maintainability and sustainability problems were addressed by an ecosystem entity. Our principles and empirically backed software architecture techniques over due course help reduce the cost of transformation and maintenance of a large mobile software ecosystem. Our new Android ecosystem model and the associated platform software architecture improvements built a foundation for fast platform software updates that are crucial for software security and mobile user experience.

## A APPENDIX

Table 6. Code Examples for .hal to .vts Conversion

**(a) .hal files***

```
interface IVehicle {
 getAllPropConfigs() generates (vec<VehiclePropConfig> propConfigs);

 getPropConfigs(vec<int32_t> props) generates (StatusCode status, vec<VehiclePropConfig> propConfigs);

 subscribe(IVehicleCallback callback, vec<SubscribeOptions> options) generates (StatusCode status);
 ...
};

interface IVehicleCallback {
 oneway onPropertyEvent(vec<VehiclePropValue> propValues);
 ...
};
```

**(b) .vts files (auto-generated)**

```
component_name: "IVehicle"
interface: {
 api: { name: "getPropConfigs"
        is_inherited: false
        arg: { name: "props"
               type: TYPE_VECTOR
               vector_value: { type: TYPE_SCALAR
                               scalar_type: "int32_t" } }
        return_type_hidl: { name: "status" ... }
        return_type_hidl: { name: "propConfigs"
                            type: TYPE_VECTOR
                            vector_value: {
                              type: TYPE_STRUCT
                               PREDEFINED_TYPE: ... } } ...
} ...
```

* **Bold** is for API names, *italic* is for keywords, *vec* is for a vector, *int32_t* is for the 32-bit signed integer type, *generates* is for synchronous return value(s), and *oneway* is for when there is no return value.

Let us use the IVehicle interface of hardware.automotive.vehicle@2.0 to explain how host-driven tests are automated in Section 4.2. As shown in Table 6(a), the vehicle@2.0 .hal files declare APIs which are implemented in the vehicle@2.0 HAL server. The HAL client implements APIs in IVehicleCallback so that the server can asynchronously return values. Host-driven tests keep the test logic in the host-side Python code that sends requests to a VTS driver on a target Android device. The VTS driver converts each request message to a C/C++ function call to a target HAL API method. Table 6(b) lists the .vts files generated and used as part of that procedure, as well as the argument and return value specifications of the getPropConfigs method. A .vts file is an ASCII protocol buffer file that keeps the intermediate representation of a converted .hal file.

Table 7(a) shows the VTS driver that is auto-generated by the VTS Compiler (vtsc). This C/C++ code is for the getPropConfigs method and shows how the driver converts a call request, which is specified in a FunctionSpecificationMessage protocol buffer message, to a C/C++ function call argument (arg0 for "props"). An HwBinder proxy object instance (hw_binder_proxy_) is used to call the respective HAL API method, which uses the function pointer provided in the second argument of hw_binder_proxy_->getPropConfigs call (see the code from [&] in Table 7(a)). The

Table 7. VTS Host-Driven Tests

**(a) VTS driver code in C/C++ (auto-generated)**

```
bool <omitted>_IVehicle::CallFunction(const FunctionSpecificationMessage& func_msg, <omitted>,
                                      FunctionSpecificationMessage* result_msg) {
const char* func_name = func_msg.name().c_str();
if (!strcmp(func_name, "getPropConfigs")) {
 hidl_vec<int32_t> arg0;
 arg0.resize(func_msg.arg(0).vector_value_size());
 for (int arg0_i = 0; arg0_i < func_msg.arg(0).vector_value_size(); arg0_i++) {
  arg0[arg0_i] = func_msg.arg(0).vector_value(arg0_i).scalar_value().int32_t();
 }
 hw_binder_proxy_->getPropConfigs(
  arg0, [&](StatusCode rtn_arg0, const hidl_vec<VehiclePropConfig>& rtn_arg1) {
   result_msg->set_name("getPropConfigs");
   VariableSpecificationMessage* rtn_v0 = result_msg->add_return_type_hidl();
   rtn_v0->set_type(TYPE_ENUM);
   SetResult_<omitted>_StatusCode(rtn_v0, rtn_arg0);
   VariableSpecificationMessage* rtn_v1 = result_msg->add_return_type_hidl();
   rtn_v1->set_type(TYPE_VECTOR);
   rtn_v1->set_vector_size(rtn_arg1.size());
   for (int rtn2_i = 0; rtn2_i < (int) rtn_arg1.size(); rtn2_i++) {
    auto* rtn_v1_vec_i = rtn_v1->add_vector_value();
    rtn_v1_vec_i->set_type(TYPE_STRUCT);
    SetResult_<omitted>_VehiclePropConfig(rtn_v1_vec_i, rtn_arg1[rtn2_idx]);
   }
 }); ...
```

**(b) Host-driven test code in Python**

```
def testApiCall(self):
 propToConfig = {}
 for config in self.vehicle.getAllPropConfigs():
  propToConfig[config['prop']] = config

def testCallback(self):
 self.onPropertyEventCalled = 0
 def onPropertyEvent(vehiclePropValues):
  self.onPropertyEventCalled += 1

 callback = self.vehicle.GetHidlCallbackInterface(
  "IVehicleCallback",
  onPropertyEvent=onPropertyEvent, <omitted>)
 self.vehicle.subscribe(callback, <omitted>)
 time.sleep(3)
 self.assertGreater(self.onPropertyEventCalled, 0)
 # test passed
```

function pointer is used to convert two synchronous return values back to VariableSpecificationMessage messages (rtn_v0 and rtn_v1 for "status" and "propConfigs", respectively).

To call a target HAL API method, the host-side test code uses an RPC mirror object (e.g., self.vehicle) and calls the getAllPropConfigs method of IVehicle. The testApiCall function shown in Table 7(b) stores the return value of the getAllPropConfigs method to a dictionary (propToConfig). This mirror object communicates with the driver by sending a message that includes FunctionSpecificationMessage and by receiving a message that includes

VariableSpecificationMessage; asynchronous callback is also supported. The testCallback function shown in Table 7(b) shows how a callback handler (onPropertyEvent) is registered using the subscribe method of the IVehicle interface. The test case checks how many times the callback handler is called on the host-side by the remote HAL running on a target device.